\documentclass[letterpaper]{article} 
\usepackage{aaai24}  
\usepackage{times}  
\usepackage{helvet}  
\usepackage{courier}  
\usepackage[hyphens]{url}  
\usepackage{graphicx} 
\usepackage{natbib}  
\usepackage{caption} 
\usepackage{algorithm}
\usepackage{algorithmic}

\usepackage{newfloat}
\usepackage{listings}

\usepackage{cite}
\usepackage{amsmath,amssymb,amsfonts}
\usepackage{algorithmic}
\usepackage{graphicx}
\usepackage{textcomp}
\usepackage{caption}

\usepackage{algorithmic}
\usepackage{graphicx}
\usepackage{textcomp}
\usepackage{textcomp}
\usepackage{verbatim}

\usepackage{xcolor}
\usepackage{placeins}
\usepackage{float}
\usepackage{subfig} 
\usepackage{subfloat}
\usepackage{cite}
\usepackage{makecell}
\usepackage{color}
\usepackage{textcomp}
\usepackage{url}
\usepackage{verbatim}
\usepackage{graphicx}
\usepackage{amsmath,amsfonts}
\usepackage{array}
\usepackage{multirow}
\usepackage{booktabs}
\usepackage{float}

\DeclareCaptionStyle{ruled}{labelfont=normalfont,labelsep=colon,strut=off} 
\floatstyle{ruled}
\newfloat{listing}{tb}{lst}{}
\floatname{listing}{Listing}
\title{Wavelet Dynamic Selection Network for Inertial Sensor Signal Enhancement}
\author{
	Yifeng Wang, Yi Zhao\thanks{Corresponding author.}
}
\affiliations{
	
	School of Science, Harbin Institute of Technology, Shenzhen\\
	wangyifeng@stu.hit.edu.cn, zhao.yi@hit.edu.cn
}

\usepackage{bibentry}

\begin{document}
	
	\maketitle

	\begin{abstract}
		As attitude and motion sensing components, inertial sensors are widely used in various portable devices, covering consumer electronics, sports health, aerospace, etc. But the severe intrinsic errors of inertial sensors heavily restrain their function implementation, especially the advanced functionality, including motion trajectory recovery and motion semantic recognition, which attracts considerable attention. As a mainstream signal processing method, wavelet is hailed as the mathematical microscope of signal due to the plentiful and diverse wavelet basis functions. However, complicated noise types and application scenarios of inertial sensors make selecting wavelet basis perplexing. To this end, we propose a wavelet dynamic selection network (WDSNet), which intelligently selects the appropriate wavelet basis for variable inertial signals. In addition, existing deep learning architectures excel at extracting features from input data but neglect to learn the characteristics of target categories, which is essential to enhance the category awareness capability, thereby improving the selection of wavelet basis. Therefore, we propose a category representation mechanism (CRM), which enables the network to extract and represent category features without increasing trainable parameters. Furthermore, CRM transforms the common fully connected network into category representations, which provide closer supervision to the feature extractor than the far and trivial one-hot classification labels. We call this process of imposing interpretability on a network and using it to supervise the feature extractor the feature supervision mechanism, and its effectiveness is demonstrated experimentally and theoretically in this paper. The enhanced inertial signal can perform impracticable tasks with regard to the original signal, such as trajectory reconstruction. Both quantitative and visual results show that WDSNet outperforms the existing methods. Remarkably, WDSNet, as a weakly-supervised method, achieves the state-of-the-art performance of all the compared fully-supervised methods.
		
	\end{abstract}

	\section{Introduction}
	
	Inertial sensors can provide information about the position, velocity, attitude, and movement patterns of a moving object in different environments \cite{shaeffer2013mems, madgwick2011estimation, esfahani2019orinet}. Due to their small size, portable wearing, and low power consumption, they are widely used in various applications such as navigation \cite{9500152}, robotics \cite{8794399}, gaming \cite{ehatisham2021expert}, and health monitoring \cite{montesinos2018wearable}. In addition, inertial sensors are not affected by light, occlusion, noise, and other external environments during data acquisition compared with optical sensors, ultrasonic sensors, and other motion capture equipment \cite{liu2020wearable, li2023attitude, chen2018ionet}.
	
	In the field of localization and navigation, inertial sensors can be used to estimate the location and trajectory of an object or a vehicle, which play a critical role when GPS signals are unavailable indoors, underground, or underwater \cite{zhuang2023multi, zhang2020navnet, ferrera2019aqualoc}. Inertial sensors can also be combined with other sensors, such as cameras, lasers, or radio signals, to improve the accuracy and robustness of localization and navigation systems \cite{qin2018vins, weber2021riann}. 
	Moreover, inertial sensors can serve to create maps of unknown environments by tracking motion \cite{esfahani2019aboldeepio}.
	In the field of health care and sports engineering, inertial sensors can be worn on the body or attached to medical devices to monitor the physical activity and health status of a person \cite{qiu2022multi}. 
	For instance, inertial sensors are employed to measure the step length, cadence, and stride variability of a Parkinson's patient \cite{nguyen2017using}, and detect falls, injuries, or abnormal behaviors of elderly people or patients \cite{tunca2019deep}.
	In the field of gaming and virtual reality, inertial sensors create immersive and interactive virtual reality experiences by tracking the motion and orientation of the user \cite{caserman2019survey, 10318077}. Inertial sensors are also used to control the movement of virtual characters by mimicking the user's actions \cite{pan2018and}. Moreover, an inertial measurement unit (IMU) can simulate realistic physical effects such as gravity, inertia, and collisions in virtual environments \cite{foehn2022agilicious}. In summary, inertial sensors are versatile and powerful for object orientation and tracking in various contexts \cite{liu2020tlio, burri2016euroc}.
	
	However, the major drawback of inertial sensors in practice is the presence of noise, including quantization noise, scale factor error, cross-coupling error, etc., which can severely affect their accuracy and reliability \cite{chen2020deep}.
	Furthermore, these errors can be accumulated in the calculation, thereby making the function realization of inertial signals much more challenging \cite{10080916, saha2022tinyodom, saha2023inertial}.
	Therefore, signal enhancement is critical to ensure the effective utilization of inertial sensors in diverse application scenes \cite{caesar2020nuscenes}. 
	
	The inertial sensor signal denoising methods can be broadly classified into model-driven and data-driven categories \cite{golestani2020human}. Data-driven methods use machine learning techniques, such as k-nearest neighbors (kNN), artificial neural networks (ANN), or convolutional neural networks (CNN) \cite{engelsman2023data} to learn the denoising function from data, without relying on explicit models of the signal.
	However, data-driven methods for IMU denoising require large amounts of labeled data for training, which may be impractical in some cases \cite{herath2020ronin}. Furthermore, deep learning methods may suffer from overfitting or generalization issues when dealing with noisy or inferior signals, especially for low-cost sensors \cite{yuan2023simple}. These defects make the data-driven methods weaker in trustworthiness and interpretability than model-driven methods.
	
	In contrast, model-driven methods can exploit the physical principles or intrinsic properties of the sensor signals \cite{9119813} to design accurate and robust denoising algorithms \cite{min2021drop}. Relying on prior knowledge of the signal or noise characteristics, Kalman filters, empirical mode decomposition, or wavelet transforms are employed to separate the noise from the signal \cite{liu2020denoising,he2019noise}. Among them, wavelet-based methods are considered the powerful and effective strategy for signal enhancement, which can adapt to signal characteristics, such as peak widths, frequency bands, or noise levels, and preserve the signal features, such as spikes, edges, or transients, in both time and frequency domains by using different wavelet functions \cite{saydjari2022equivariant}. However, the accuracy and efficiency of the wavelet-based methods rely on their wavelet basis functions, which have specification characteristics, including filter length, symmetry, vanishing moment, regularity, and similarity, which determine their suitability to different types of signals. Therefore, it is essential to select wavelet basis functions in the wavelet implementation for signal enhancement.
	
	Currently, although wavelet-based methods are widely used in the field of signal processing, the selection of wavelet basis is subjective or empirical, as there is no universal criterion or rule to choose an appropriate wavelet basis function for a given signal. To this end, we propose a wavelet dynamic selection network (WDSNet), which intelligently selects the appropriate wavelet basis according to the characteristics of the input signal for the best signal enhancement. Specifically, we propose a category representation mechanism that acts on the fully connected (FC) layer, making each column of its weight matrix encodes a category (i.e., a candidate wavelet). In this way, the characteristics of each wavelet are memorized by the network, so the WDSNet can find the most suitable wavelet based on the feature of the input data. The contributions of this paper are as follows:
	
	\begin{itemize}
		\item This paper proposes a universal wavelet basis selection paradigm, WDSNet, to achieve inertial sensor signal quality enhancement for the first time. The WDSNet has the advantages of not only the flexibility of data-driven methods but also the reliability and interpretability of model-driven methods.
		\item This paper proposes a category representation mechanism (CRM) that can improve the category awareness ability by constructing category features in the classifier without extra training parameters, thereby capturing the properties of candidate wavelets for suitable selection. 
		\item This paper proposes a feature supervision mechanism (FSM) that uses the network parameters to directly supervise feature extraction by giving the net interpretability. Specifically, the designed CRM turns a fully connected net into a category representation dictionary, providing closer supervision than the far class labels, thus alleviating gradient vanishing, which is verified experimentally and theoretically in this paper.
		
		\item We comprehensively compare the existing IMU signal enhancement methods in terms of Allan variance analysis and four downstream tasks.
		The experimental results show that the proposed WDSNet achieves state-of-the-art performance and has significant superiority compared with all current IMU signal enhancement methods.
	\end{itemize}

	\section{Related works}
	\subsection{Model-Driven Methods for IMU Signal Denoising}
	Model-driven methods for IMU denoising aim to exploit the physical properties of the inertial sensors and the motion dynamics to reduce the noise. One common approach was to use various filters to correct the signal \cite{9198927}. However, it required accurate sensor performance parameters and noise distribution, which are often difficult to obtain in practice \cite{toft2020long}.
	Another common approach was the utilization of wavelet-based methods to enhance IMU signal quality. For example, the wavelet thresholding method was widely used to remove the high-frequency noise in the IMU data and preserve the low-frequency signals \cite{wu2019survey}, which improved the accuracy of attitude estimation and navigation. However, current wavelet-based methods rely on manually selecting a wavelet basis, which may not be optimal for variable noise types and levels.

	\subsection{Data-Driven Methods for IMU Signal Denoising}
	Data-driven methods for enhancing the signal quality of inertial sensors have attracted considerable attention in recent years due to the increasing demand for accurate and robust inertial sensing.
	Chen \textit{et al.} \cite{chen2022towards} assumed that IMU signals could be approximated as constants in a short time. Based on this assumption, they divided an IMU signal into multiple short segments, each segment being labeled with a constant as the ideal value, and then used a CNN to predict the ideal value based on these segments. However, this would result in the enhanced signals changing from continuous signals to square wave pulse signals. Moreover, since all the training samples only covered linear and circular motions, this signal enhancement method became invalid under the scenario of complicated or arbitrary motion. Han \textit{et al.} \cite{han2021hybrid} fixed the IMU on a high-precision turntable, and collected gyroscope data and real angular velocity data at the same time. Using the ideal data, they built a hybrid deep recurrent neural network (GRU-LSTM), which implemented end-to-end supervised training. This method achieved good signal enhancement results on the training set, but the signal enhancement results on the test data were unsatisfactory. Further, Boronakhin \textit{et al.} \cite{boronakhin2022optimization} used Bayesian optimization to optimize the hyperparameters of the GRU-LSTM model, which improved the model performance on the test set, but due to the lack of interpretability of deep learning models, the generated signal has poor trustworthiness and even loses the semantic information of the input signals \cite{shamwell2019unsupervised}. 
	Even though the generated signals were better than the input signals in some quantization indicators (such as quantization noise, angle random walk, velocity random walk, and bias instability), they performed worse than the original signals in downstream tasks such as attitude estimation and trajectory reconstruction \cite{huang2019apolloscape}. In summary, due to the mechanism that the deep learning method achieves signal enhancement in a generative paradigm, the signal enhancement process is uncontrollable. Albeit the generated IMU signals have improvements in quantization indicators, they are practically unusable due to the lack of authenticity and reliability.

	\section{Methodology}

	\subsection{Framework Overview}
	\begin{figure*}[htbp]
		\centering
		\includegraphics[width=1\linewidth]{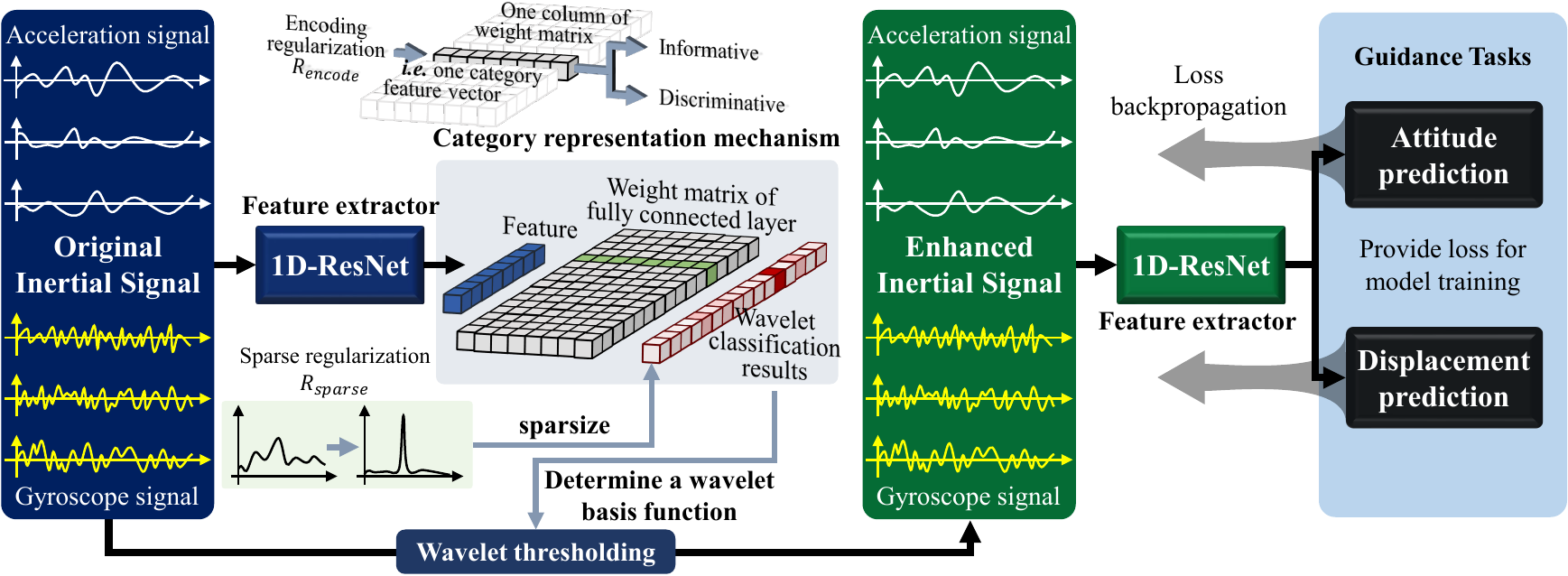}
		\caption{An overview of the WDSNet framework. The input data is fed into a 1D-ResNet for feature extraction, and an FC layer serves as a classifier that performs wavelet selection based on the extracted features. Then the selected wavelet is used to perform wavelet thresholding denoising on the original signal. The enhanced signal is input into another 1D-ResNet for attitude prediction and displacement prediction, which supervises the wavelet selection through the backpropagation of task losses.}
		\label{Framework}
	\end{figure*}
	
	Wavelet can extract local spectral and temporal information simultaneously, capture the transient features of signals, and is capable of multi-resolution analysis. It is a powerful signal enhancement tool that meets various denoising requirements \cite{ma2020end}. In this process, the selection of wavelets is crucial. Ideally, the selected wavelet should satisfy orthogonality, high vanishing moments, compact support, and symmetry/anti-symmetry. Practically, such a wavelet does not exist because only the Haar wavelet is symmetric/anti-symmetric, while high vanishing moments and compact support are a pair of contradictory quantities \cite{Wang2020AdaptiveSK}. Therefore, selecting the appropriate wavelet according to input data is an essential but unsolved problem. To address this problem, we design a wavelet dynamic selection network (WDSNet) for signal quality enhancement, which encodes the features of 16 wavelet bases in the network, thereby achieving the most appropriate wavelet selection.
	The framework of WDSNet is presented in Fig. \ref{Framework}. 
	
	Firstly, a 1D-ResNet is set to extract the input data features, and then an FC layer performs classification (i.e., wavelet selection) based on the extracted features, where each category corresponds to a wavelet. We use the selected wavelet basis to perform wavelet thresholding on the input signal and then obtain the enhanced signal. To supervise the wavelet selection process, we force the enhanced signal to complete two guidance tasks, i.e., attitude prediction and displacement prediction. The smaller error of the guidance tasks, the higher quality of the enhanced signal, indicating that the wavelet selection is more appropriate. In such a way, these two tasks jointly guide the network to select the most suitable wavelet for signal processing. It is worth noting that compared with the fully supervised deep learning-based signal enhancement methods, the guidance tasks require much less supervision information, which is convenient to deploy.

	\subsection{Category Representation Mechanism}
	Although the deep learning classifier excels at mining and extracting input data features, it neglects to capture the underlying features of target categories, which limits the performance of the classifier. Specifically, the lack of awareness for wavelet bases can lead to an inappropriate selection of wavelets.
	To address this issue, we propose a category representation mechanism (CRM), which comprises a sparsity regularization term $R_{sparse}$ and an encoding regularization term $R_{encode}$, where $R_{sparse}$ forces the weight matrix of the FC layer to represent category features and $R_{encode}$ forces category features to be more discriminative and informative.
	
	The sparsity regularization term acts on the output layer ${R_{sparse}} = {\left\| {\hat y} \right\|_1}$, to make the classification vector $\hat{y}$ sparse, which means that only one element in this vector is close to 1 and the rest are close to 0. 
	Since the elements close to 0 in $\hat{y}$ have negligible effect in the subsequent calculations, the backward propagation of loss is truncated here, and the corresponding columns in the weight matrix are not updated, which ensures that each element in the output vector matches exactly each column in the weight matrix.

	Under the influence of sparse regularization terms, the loss of each wavelet selection is only related to a specific column of the weight matrix, which can be regarded as the vector representation of the corresponding wavelet category. The ideal vector representation of wavelet categories should be discriminative and informative. Specifically, category representation vectors should be orthogonal to each other for reflecting the diversity between different categories as much as possible. Furthermore, the vectors need to store as much information as possible rather than being as trivial as one-hot encoding.
	We, therefore, design an encoding regularization term that makes category vectors more orthogonal and informative. Its specific operation is as follows.
	
	We first use the $\alpha$-order Rényi information entropy $S_\alpha$ to measure the amount of information in the weight matrix $W$, which is calculated by the following equations:
	\begin{equation}
		\label{S_alpha}
		{S_\alpha }(W) = \frac{1}{{1 - \alpha }}{\log _2}(tr({\tilde G^\alpha })),
	\end{equation}
	\begin{equation}
		\label{G tilde}
		\tilde G[i][j] = \frac{1}{{16}}\frac{{G[i][j]}}{{\sqrt {G[i][i] \cdot G[j][j]} }},
	\end{equation}
	\begin{equation}
		\label{G}
		G[i][j] = \left\langle {{W^{(i)}},{W^{(j)}}} \right\rangle,
	\end{equation}
	where $G$ denotes the Gram matrix of the weight matrix $W$, $G[i][j]$ is the inner product of the $i$-th column and $j$-th column of $W$, and $\tilde G$ is the trace-normalized $G$, i.e., $tr(\tilde G)=1$. $\alpha$ is set to 2 according to \cite{8787866, 8998186}.
	${S_\alpha}(W)$ measures the amount of information in the category matrix (weight matrix, as shown in the grey blocks in Fig. \ref{Framework}). The larger ${S_\alpha}(W)$ is, the more information the matrix $W$ has, indicating that the category representation vectors are more informative. Simultaneously, as ${S_\alpha}(W)$ increases, all elements in the Gram matrix $G$ are forced to descend, thereby indicating that the inner product of different class encoding vectors is smaller and the vectors have stronger orthogonality. Hence, the two requirements for category representation vectors can be satisfied as ${S_\alpha}(W)$ increases. Therefore, we propose a regularization term $R_{encode}$ in equation \ref{R_encode}. As this regularization term decreases with training, ${S_\alpha}(W)$ gradually increases so we can obtain a discriminative and informative vector representation for each class.
	\begin{equation}
		\label{R_encode}
		{R_{encode}} = \frac{1}{{{S_\alpha}(W)}}.
	\end{equation}
	CRM transforms the weakly interpretable FC layer into the encodings of the target categories, which enables the weight matrix to be informative and thus avoids overfitting.

	\subsection{Feature Supervision Mechanism}
	CRM enhances the perception ability for target categorization by extracting target category features. We found that these category features can provide direct supervision for deep learning networks. 
	For a general deep learning network, the input data $x$ is fed into the feature extractor $f(\cdot)$ to obtain the data feature $h=f(x)$, then the feature $h$ is fed into a task head $g(\cdot)$ for the final prediction results $\hat y = g(h)$.
	The supervision information $y$ acts on the results $\hat y$, resulting in a loss function $L(\hat y,y)$, which is back-propagated to guide the weight updating of feature extractor $f(\cdot)$ and task head $g(\cdot)$. 
	As the network gradually deepens and becomes more complex, the shallow structure, such as the feature extractor, may not be effectively supervised by the far classification labels due to gradient vanishing, gradient explosion, or other unknown reasons.

\begin{figure}[!h]
		\centering
		\includegraphics[width=\linewidth]{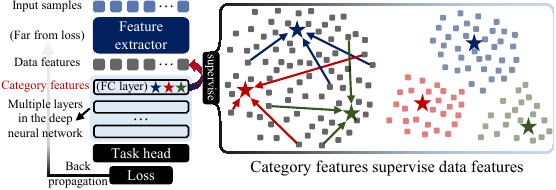}
		\caption{Illustration of feature supervision mechanism. Pentagrams represent category features, and small blocks represent data features. Under the supervision of category features, data features exhibit category characteristics and become more discriminative.}
		\label{FSM Illustration}
	\end{figure}

	The proposed CRM provides a feasible solution to this problem by internalizing category representation in the network, as defined in equation \ref{maxS}.
	\begin{equation}
		\label{maxS}
		\begin{array}{l}
			{g_{CRM}}(h) = {{\hat y}_{CRM}}\\
			\min L({{\hat y}_{CRM}},y) \Rightarrow \max \left\langle {h,W_{CRM}^{(n)}} \right\rangle 
		\end{array},
	\end{equation}
	where $g_{CRM}(\cdot)$ and ${\hat y}_{CRM}$ represent the task head and output results after applying CRM, $h$ is still the extracted feature, and $W_{CRM}^{(n)}$ represents the $n$-th column weight in the classifier, i.e., the feature vector of the $n$-th category. $\left\langle { \cdot , \cdot } \right\rangle $ represents the inner product, which reflects the similarity between the two vectors. It can be observed that accurate classification requires that the data feature $h$ has a high similarity with the target category feature $W_{CRM}^{(n)}$. Therefore, the supervision of outputs is transformed into the supervision of the features, then transformed into the supervision of the feature extractor $f(\cdot)$. This process is named a feature supervision mechanism (FSM), which transforms the fully connected (FC) layer closest to the data features into a category feature layer, allowing it to directly supervise the data features, thus alleviating the gradient vanishing caused by the excessive neural network layers, as shown in Fig. \ref{FSM Illustration}.

	\subsection{Weakly-Supervised Guidance Tasks}
	Obtaining frame-paired annotations for end-to-end training in the IMU signal enhancement task is extremely challenging \cite{marchetti2020multiple}. To address this issue, we set up two tasks with readily available labels as guidance tasks, displacement prediction, and attitude prediction, to guide the selection of the wavelet. 
	
	IMU provides acceleration and angular velocity data during motion as a motion sensor. The attitude prediction task requires the enhanced signal to predict the changes of three attitude angles: yaw, roll, and pitch. The displacement prediction task requires the enhanced signal to predict the displacement vector corresponding to the motion. These two tasks assess the motion information of the enhanced signal in terms of spatial position change and attitude rotation change, so they are configured to guide the improvement of signals.
	
	\section{Experiments and Results}
	\subsection{Experiment Dataset}
	The built-in IMUs are the most widely used and representative low-cost inertial sensors. We take 15 smartphones with the built-in IMUs to collect the inertial dataset, of which one type of smartphone is employed for collecting the training set, while the data of all the other phones are used for testing. The types of smartphones and their internal IMU specifications are shown in Table \ref{IMU specification}. 
	\begin{table}[!h]
		\centering
		
		\resizebox{\linewidth}{!}{
			\begin{tabular}{c|c|c|c}
				\toprule
				\multicolumn{1}{c|}{Dataset} & Smartphone & \multicolumn{1}{c|}{IMU} & \multicolumn{1}{p{4.19em}}{Unit price} \\
				\midrule
				\multicolumn{1}{c|}{Training} & HUAWEI Mate30 Pro &ICM20690	&\$0.28 \\
				\midrule
				\multicolumn{1}{c|}{\multirow{14}[20]{*}{Testing}} & HUAWEI P40 &LSM6DSM	&\$0.30 \\
				\cmidrule{2-4}          & HUAWEI P40 Pro &LSM6DSO	&\$0.33 \\
				\cmidrule{2-4}          & iPhone 7 Plus &ICM20600	&\$0.20 \\
				\cmidrule{2-4}          & SAMSUNG Galaxy S7 &LSM6DS3	&\$0.20 \\
				\cmidrule{2-4}          & SAMSUNG Galaxy S8 &LSM6DSL	&\$0.26 \\
				\cmidrule{2-4}          & Realme GT &BMI160 	&\$0.21 \\
				\cmidrule{2-4}          & Xiaomi 11 &BHI260AB	&\$0.30 \\
				\cmidrule{2-4}          & OPPO Reno 6 &ICM-40607	&\$0.28 \\
				\cmidrule{2-4}          & Lenovo Legion Phone &ICM-42605	&\$0.20  \\
				\cmidrule{2-4}          & VIVO X30 &LSM6DSM	&\$0.30  \\
				\cmidrule{2-4}          & VIVO T2x &LSM6DSO	&\$0.33	\\
				\cmidrule{2-4}          & iPhone 13 &Undisclosed	&/  \\
				\cmidrule{2-4}          & iPhone 12 &Undisclosed	&/  \\
				\cmidrule{2-4}          & iPhone 11 Pro &Undisclosed	&/  \\
				\bottomrule
			\end{tabular}%
		}
		\caption{The built-in IMU specifications of some smartphones. Note that since the IMUs in some types of iPhones are customized by the manufacturer, the model and price are not disclosed.}
		\label{IMU specification}%
	\end{table}%
	It can be found that the price of inertial sensors used in our experiments does not exceed \$0.5.
	Meanwhile, the phone is fixed at the flange of the mechanical arm (ROKAE xMate ER3 Pro), which is employed to accurately record the changes of attitude and position as the labels of the previous two guidance tasks.
	All experiments are implemented by Pytorch 1.10.1 with an Nvidia RTX 2080TI GPU and Intel(R) Xeon(R) W-2133 CPU.
	
	\subsection{Comparative Results}
	\subsubsection{Static Evaluation}
	Allan variance is a classical time-domain technique that provides the quantitative indicators of IMU signal quality, including quantization noise (QN), angle random walk (ARW), velocity random walk (VRW), and bias instability (BI).
	Based on these indicators, we compared WDSNet with the popular IMU signal enhancement methods in recent years, and the results are reported in Table \ref{allen comparison}.
	All the comparative methods are implemented strictly following the conditions in their papers or using their open-source codes.
	Some data-driven methods perform relatively poorly since they act as a black box to generate enhanced signals, resulting in unreliable signal enhancement. In contrast, model-driven methods optimize the signal from an interpretative perspective, making the enhanced signal more favorable. However, model-driven methods cannot make flexible adjustments according to the characteristics of the input signal, thus presenting a bottleneck in signal quality enhancement.
	By combining model-driven and data-driven strategies, WDSNet achieves the best performance.

	\begin{table}[htbp]
		\centering
		\setlength{\tabcolsep}{1.2mm}
		{
		\resizebox{\linewidth}{!}{
			\begin{tabular}{c|c|c|c|c|c|c|c}
				\toprule
				\multicolumn{2}{c|}{\multirow{2}[2]{*}{Architecture}} & \multicolumn{3}{c|}{Acceleration} & \multicolumn{3}{c}{Angular Velocity} \\
				\cmidrule{3-8}    \multicolumn{2}{c|}{} & \multicolumn{1}{c|}{QN} & \multicolumn{1}{c|}{VRW} & \multicolumn{1}{c|}{BI} & \multicolumn{1}{c|}{QN} & \multicolumn{1}{c|}{ARW} & \multicolumn{1}{c}{BI} \\
				\midrule
				\multicolumn{2}{c|}{Raw signal} & 1.21 & 1.92 & 3.12 & 3.55 & 5.98 & 9.06 \\
				\midrule
				\multirow{2}[2]{*}{\makecell[c]{Model\\Driven}} 
				& Kalman filter
				& 0.49 & 0.58 & 0.64 & 0.77 & 0.85 & 1.40 \\
				\cmidrule{2-8}    
				\multicolumn{1}{c|}{} & SG filter
				& 0.54 & 0.71 & 0.76 & 0.98 & 1.10 & 1.74 \\
				\midrule
				\multirow{4}[5]{*}{\makecell[c]{Data\\Driven}} & CNN
				& 0.56 & 0.70 & 0.77 & 1.06 & 1.08 & 1.64 \\
				\cmidrule{2-8}    
				\multicolumn{1}{c|}{} & GRU-LSTM
				& 0.52 & 0.60 & 0.71 & 0.77 & 0.96 & 1.53 \\
				\cmidrule{2-8}    
				\multicolumn{1}{c|}{} & Opt-GRU-LSTM
				& 0.40 & 0.53 & 0.61 & 0.65 & 0.72 & 0.98 \\
				\cmidrule{2-8}    
				\multicolumn{1}{c|}{} & kNN
				& \underline{0.20} & \underline{0.31} & \underline{0.47} & \underline{0.48} & \underline{0.55} & \underline{0.65} \\
				\midrule
				\makecell[c]{Model\\-Data} & WDSNet (Ours) 
				& \textbf{0.06} & \textbf{0.07} & \textbf{0.08} & \textbf{0.09} & \textbf{0.09} & \textbf{0.13} \\
				\bottomrule
			\end{tabular}%
		}
		\caption{Performance comparison of mainstream methods in terms of Allan variance analysis. We bold the best and underline the suboptimal results.}
		\label{allen comparison}%
	}
	\end{table}%

	\subsubsection{Dynamic Evaluation}
		\begin{table*}[htbp]
		\centering
		\resizebox{\linewidth}{!}{
			\begin{tabular}{c|c|c|c|c|c}
				\toprule
				\multicolumn{2}{c|}{Architecture} & \multicolumn{1}{c|}{\makecell[c]{Error of attitude \\ estimation (deg) $\downarrow$}} & \multicolumn{1}{c|}{\makecell[c]{Error of position \\ estimation (m) $\downarrow$}} & \multicolumn{1}{c|}{\makecell[c]{Accuracy of semantic \\ recognition $\uparrow$}} & \multicolumn{1}{c}{\makecell[c]{FSSE of trajectory \\ reconstruction $\downarrow$}} \\
				\midrule
				\multicolumn{2}{c|}{Raw signal (No processing)} &  10.69 & 1.09 & 78.42\% & 0.78 \\
				\midrule
				\multirow{2}[2]{*}{Model Driven} & EMD-Kalman filter 
				& 5.36 (-49.83\%)  & \underline{0.38 (-65.27\%)} & 85.28\% (+8.74\%) & \underline{0.18 (-76.53\%)} \\
				\cmidrule{2-6}    \multicolumn{1}{c|}{} & Savitzky Golay filter 
				& 6.27 (-41.29\%)  & 0.43 (-59.82\%) & 84.99\% (+8.38\%) & 0.22 (-72.42\%) \\
				\midrule
				\multirow{4}[5]{*}{Data Driven} & CNN 
				& 5.85 (-45.31\%)  & 0.48 (-56.13\%) & 85.52\% (+9.05\%) & 0.29 (-62.24\%) \\
				\cmidrule{2-6}    \multicolumn{1}{c|}{} & GRU-LSTM 
				& 4.94 (-53.84\%)  & 0.41 (-62.59\%) & 86.13\% (+9.83\%) & 0.27 (-65.17\%) \\
				\cmidrule{2-6}    \multicolumn{1}{c|}{} & Optimized GRU-LSTM 
				& \underline{4.43 (-58.53\%)}  & 0.39 (-64.82\%) & \underline{86.22\% (+9.95\%)} & 0.25 (-68.39\%) \\
				\cmidrule{2-6}    \multicolumn{1}{c|}{} & kNN   
				& 6.47 (-39.46\%)  & 0.46 (-57.94\%) & 85.62\% (+9.18\%) & 0.28 (-63.72\%) \\
				\midrule
				Model-Data Driven & WDSNet (Ours) 
				& \textbf{3.79 (-64.35\%)}  & \textbf{0.28 (-73.62\%)} & \textbf{87.01\% (+10.95\%)} & \textbf{0.10 (-86.95\%)} \\
				\bottomrule
			\end{tabular}%
		}
		\caption{Performance comparison of the proposed method and typical methods for  four downstream tasks. Considering the units and value ranges of four downstream tasks, we give the improvement compared with the raw signal in parentheses for convenient comparison. Note that we bold the best and underline the suboptimal results.}
		\label{task comparison}%
	\end{table*}%

		\begin{table*}[!h]
	\centering
	\resizebox{\linewidth}{!}{
		\begin{tabular}{c|c|c|c|c|c|c|c|c|c|c|c}
			\toprule
			\multicolumn{2}{c|}{Architecture} & \multicolumn{3}{c|}{Acceleration} & \multicolumn{3}{c|}{Angular Velocity} & \multicolumn{1}{c|}{\multirow{2}[5]{*}{\makecell[c]{Attitude \\ estimation \\ error (deg) $\downarrow$}}} & \multicolumn{1}{c|}{\multirow{2}[5]{*}{\makecell[c]{Position \\ estimation \\ error (m) $\downarrow$ }}} & \multicolumn{1}{c|}{\multirow{2}[5]{*}{\makecell[c]{Semantic \\recognition\\accuracy $\uparrow$}}} & \multicolumn{1}{c}{\multirow{2}[5]{*}{\makecell[c]{Trajectory\\ reconstruction\\error $\downarrow$}}} \\
			\cmidrule{1-8}    \multicolumn{1}{c|}{CRM} & \multicolumn{1}{c|}{\makecell{wavelet \\number}} & \multicolumn{1}{c|}{QN} & \multicolumn{1}{c|}{VRW} & \multicolumn{1}{c|}{BI} & \multicolumn{1}{c|}{QN} & \multicolumn{1}{c|}{ARW} & \multicolumn{1}{c|}{BI} &       &       &       &  \\
			\midrule
			\multicolumn{1}{c|}{\multirow{3}[3]{*}{\makecell[c]{w/o\\CRM}}} & 5     & 0.46 & 0.74 & 0.90 & 0.94 & 1.09 & 1.70 & 6.24 & 0.49  & 84.99\% & 0.28 \\
			\cmidrule{2-12}          & 10    & 0.39 & 0.67 & 0.71 & 0.89 & 0.93 & 1.54 & 5.91 & 0.43 & 85.55\% & 0.24 \\
			\cmidrule{2-12}          & 16    & 0.40 & 0.74 & 0.87 & 0.95 & 0.98 & 1.55 & 6.09 & 0.48 & 85.21\% & 0.28 \\
			\midrule
			\multicolumn{1}{c|}{\multirow{3}[3]{*}{\makecell[c]{w/\\CRM}}} & 5     & 0.42 & 0.70 & 0.73 & 0.74 & 0.92 & 1.60 & 6.18 & 0.41 & 85.53\% & 0.25 \\
			\cmidrule{2-12}          & 10    & \underline{0.19} & \underline{0.21} & \underline{0.26} & \underline{0.32} & \underline{0.40} & \underline{0.84} & \underline{4.71} & \underline{0.35} & \underline{86.16\%} & \underline{0.20} \\
			\cmidrule{2-12}          & 16    & \textbf{0.06} & \textbf{0.07} & \textbf{0.08} & \textbf{0.09} & \textbf{0.09} & \textbf{0.13} & \textbf{3.79}  & \textbf{0.28} & \textbf{87.01\%} & \textbf{0.10} \\
			\bottomrule
		\end{tabular}%
	}
	\caption{Ablation experiments on candidate wavelet number and CRM. The improvement on four downstream tasks compared with the raw signal is given in parentheses for convenient comparison. We bold the best and underline the suboptimal results.}
	\label{ablation}%
\end{table*}%

	Allan variance analysis is usually used to evaluate the noise of IMU signals under static conditions, and it is not adequate to evaluate the signal quality enhancement in motion. Therefore, we design four downstream tasks to examine the effectiveness of different methods. Specifically, the four downstream tasks are attitude change estimation, position change estimation (i.e., displacement estimation), motion semantic recognition, and motion trajectory reconstruction. The classic navigation algorithm is employed to perform motion trajectory reconstruction, and a ResNet containing 8 bottleneck blocks is constructed for motion semantic recognition.

	Considering the issue of inconsistent coordinate systems between the reconstructed trajectory and the real trajectory recorded by the robotic arm, we use the Fréchet spline sliding error (FSSE) \cite{10080916} to calculate the error in the trajectory reconstruction task, which can measure the morphological similarity of two spatial curves even if they are in different coordinate systems. 
	The performance comparison for the four downstream tasks is presented in Table \ref{task comparison}.
	It can be observed that the proposed WDSNet still achieves the best performance. In contrast, the kNN method, which performs well in Allan variance analysis, does not serve downstream tasks well. We note that signal enhancement has limited assistance in semantic recognition since ill-writing motion and individual writing habits, instead of signal noise, bring about the main challenges for the semantic recognition task.

	\subsection{Ablation Study}
	\begin{figure*}[!h]
		\centering
		\includegraphics[width=1\linewidth]{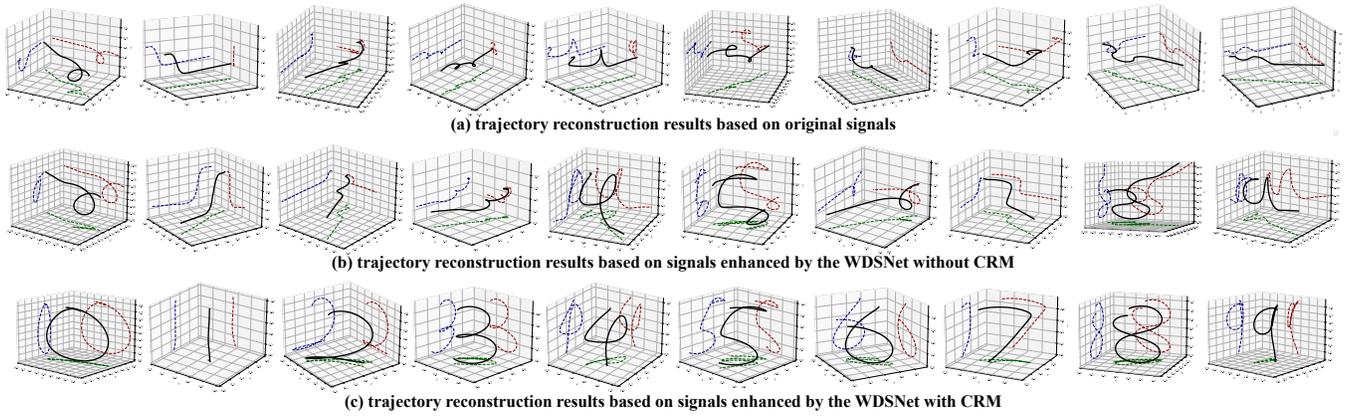}
		\caption{Trajectory reconstruction visualization for ablation study. The dashed lines in each panel indicate the projection of a reconstructed trajectory in the XY, XZ, and YZ planes.}
		\label{Trajectory}
	\end{figure*}
	\begin{figure}[htbp]
	\centering
	\includegraphics[width=1.00\linewidth]{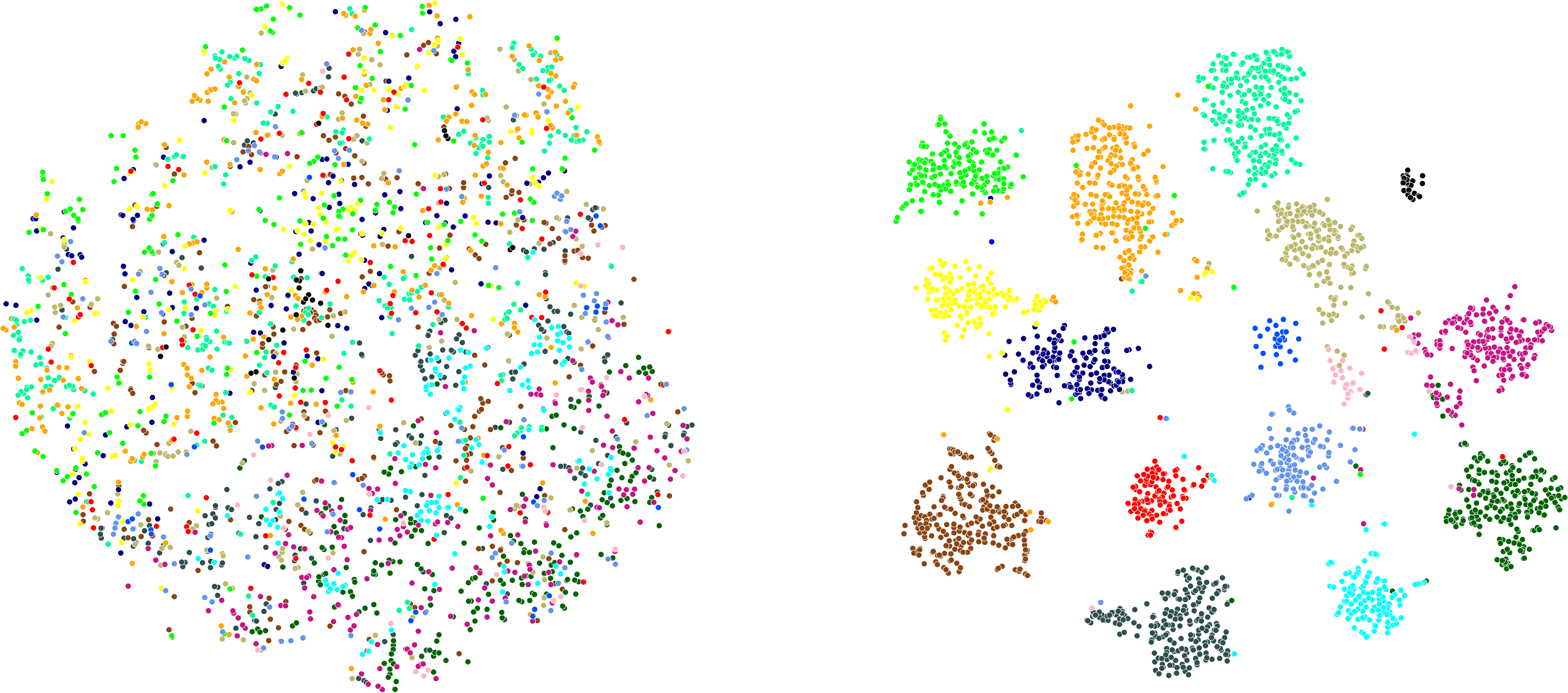}
	\caption{Effect visualization of with (right panel) and without (left panel) FSM. Each point represents the feature of a sample, and 16 colors correspond to the 16 wavelets. The data features are clustered according to wavelet categories under category feature supervision.}
	\label{FSM ablation}
    \end{figure}

	We then conduct ablation experiments to verify the effect of CRM and the number of candidate wavelet bases. The related static and dynamic evaluation results are summarized in Table \ref{ablation}.
	In comparison with the results of the existing methods in Tables 2 and 3, our wavelet selection framework without CRM still achieves significant signal enhancement. However, as more candidate wavelets are available for selection, the model is perplexed by massive candidate wavelets to determine the suitable one.
	Under this scenario, the selected wavelet often is not suitable to adapt to the input signal, such that the performance of signal enhancement does not increase but decreases.
	The introduction of CRM enhances the wavelet perception ability of the model, thereby enabling it to select the most suitable wavelet from numerous candidates according to signal properties and achieve the best signal enhancement.
	To visualize the signal enhancement of the proposed method, we reconstruct the concrete trajectories of the handwriting signals. It can be observed from Fig. \ref{Trajectory} (a) that the trajectory reconstruction based on the original signal is deplorable, and the written character is almost invisible. Fig. \ref{Trajectory} (b) illustrates that the signals enhanced by WDSNet (without CRM) are conducive to improving trajectory reconstruction. However, the lack of CRM leads to inaccurate wavelet selection, so most reconstructed trajectories are not visible enough to identify their motion characters. The model with CRM can allocate the appropriate wavelet basis for arbitrary input signals, so the improved signal can achieve much better trajectory reconstruction, as shown in Fig. \ref{Trajectory} (c).

	\subsection{Visualization Analysis on the Effect of FSM}

	To verify the effect of FSM, we utilize the t-SNE technique to perform dimensionality reduction on the features extracted by 1D-ResNet, and then compare the data feature distributions with and without category feature supervision, as shown in Fig. \ref{FSM ablation}. The left panel presents the distribution of data features supervised by loss backpropagation from the result layer, while the right one presents the distribution of data features directly supervised by category features represented within the network. It can be observed that under the FSM, the extracted data features become well-organized from stochastic distribution. Specifically, the features of the same category have stronger similarity, indicating that the model can extract more discriminative features from the input data, thereby achieving accurate classification.

	\section{Conclusion}
	In this paper, we propose a signal enhancement method, wavelet dynamic selection network (WDSNet), combining wavelet with deep learning, which holds both the reliability of model-driven approaches and the flexibility of data-driven approaches.
	To select an appropriate wavelet for signal enhancement, we propose a category representation mechanism (CRM) that improves the awareness of wavelet characteristics by learning their category representation. As a plug-and-play module, CRM improves the classification ability of deep learning without increasing trainable parameters.
	In addition, since the CRM constructs vector representations of target categories within the network, the feature extraction can be directly supervised by these category vectors. This feature supervision mechanism (FSM) is more direct and efficient than the loss backpropagation from the result of the far output layer, which has been verified theoretically and experimentally in this paper.
	Specifically, the Allan variance analysis is employed to evaluate the signal enhancement effect compared with existing methods, and we also construct four downstream tasks (posture prediction, position prediction, semantic recognition, and trajectory reconstruction) to evaluate the practical effects of enhanced signals. The results show that WDSNet achieves SOTA performance in all comparative experiments. Importantly, signals enhanced by our WDSNet perform satisfactory reconstruction of arbitrary spatial trajectories, which is usually considered an impossible function for a low-cost inertial sensor.
	On the other hand, this paper provides a universal wavelet selection strategy without requirements of any selection labels, which expands wavelet methods to be qualified for complex and ever-changing application scenarios.
	
	\appendix
	
	
	\bibliography{reference.bib}
	
\end{document}